\documentclass[preprint2]{aastex631}

\usepackage{ulem}

\newcommand{\TD}{{\tt TD}~}

\newcommand{\species}[2]{{\hbox{#1\,{\sc #2}}~}}

\shorttitle{ISM in EoR Galaxies}
\shortauthors{Ku\v{s}mi\'{c} et al.}
\graphicspath{{./}{figures/}}

\begin{document}

\title{Understanding Interstellar Metals during Reionization with Radiative SPH Simulation: Metallicity and Emission Lines from the ISM at $10 \geq z \geq 5$}

\correspondingauthor{Samir Ku\v{s}mi\'{c}}
\email{samirkusmic@gmail.com}

\author[0000-0002-0761-1985]{Samir Ku\v{s}mi\'{c}}
\affiliation{Independent Researcher}
\affiliation{New Mexico State University \\
MSC 4500, PO BOX 30001 \\
Las Cruces, NM 88003}

\author[0000-0002-0496-1656]{Kristian Finlator}
\affiliation{New Mexico State University \\
MSC 4500, PO BOX 30001 \\
Las Cruces, NM 88003}
\affiliation{Cosmic Dawn Center (DAWN) \\
Niels Bohr Institute, University of Copenhagen / DTU-Space \\ 
Technical University of Denmark}

\author[0009-0004-8503-0483]{Ezra Huscher}
\affiliation{New Mexico State University \\
MSC 4500, PO BOX 30001 \\
Las Cruces, NM 88003}

\author[0009-0002-0435-5055]{Maya Steen}
\affiliation{New Mexico State University \\
MSC 4500, PO BOX 30001 \\
Las Cruces, NM 88003}




\begin{abstract}
We compare the \texttt{Technicolor Dawn} cosmological simulations with recent observations of galactic nebular line emission during the Epoch of Reionization, providing stringent tests of the predicted ionization and metal enrichment levels. We validate the simulated population with the UVLF and $M_{\mathrm{UV}}-M_*$ relation and see that the simulated results are consistent with observations at lower masses. We extract local gas volumetric grids of density and mass-weighted metallicity, then we use \texttt{Cloudy} to produce synthetic emission spectra of \species{H}{ii} regions. The mass-metallicity relation does not evolve, which is also consistent with observations. The predicted oxygen abundance exceeds observational inferences by about 0.5 dex, suggesting either overly efficient enrichment or weak feedback. However, applying the O32 diagnostic directly to our synthetic spectra shows an offset of 1 dex from the correct outputted gas-phase metallicity. This suggests that O32 is biased high at a level that is more than sufficient to account for the simulation-observation offset. The simulated galaxies' line diagnostics show mostly weaker [\species{O}{iii}] lines and lower diagnostic values of O3 and Ne3O2 compared to observations. This suggests higher ionization parameters within the simulated galactic population in general.

\end{abstract}


\newcommand{\ez}[1]{\textcolor{teal}{#1}}

\section{Introduction} \label{sec:intro}


The coupled and poorly understood processes governing galaxy evolution leave long-lived imprints in the interstellar medium (ISM) that can be interpreted through comparison with suitably realistic models. Their imprints may persist into the local Universe such as with globular clusters, but direct observations of early times provide immediate snapshots at or near the beginning and therefore the actual formation and feedback at that time. The James Webb Space Telescope (JWST) enables inquiries into the internal nature of representative galaxy samples during the Epoch of Reionization (EoR) through large surveys such as UNCOVER and CEERS \citep[][respectively]{Price:2024, Finkelstein:2023}. These studies are the most direct method of examining the first stages of galactic evolution, the history of our Universe, the start of the baryon cycle \citep[][]{Tumlinson:2017}{}{}, and the history of reionization \citep{Stark:2016}.



The earliest stages of galaxy evolution were regulated primarily by feedback from massive stars, whose  byproducts include metals and ionizing photons. These manifest observationally within the ISM through metallicity $Z$ and ionization parameter $U$ or $Q = U c$. Line diagnostics are used, such as $\mathrm{O3} = [\species{O}{iii}]5007/ \species{H}{$\beta$} $ for metallicity, or $\mathrm{Ne3O2} = [\species{Ne}{iii}]3870/[\species{O}{ii}]3728$ for ionization \citep[][]{Hu:2024, Trump:2023, Cameron:2023, Tang:2023, Nakajima:2023}{}{}, both of which can be detected readily by the JWST. 

New observational constraints allow model building to improve our understanding of star formation, feedback, and ionizing photon production within $\species{H}{ii}$ regions. Their ionization parameter is related to the ionizing photon production efficiency and escape fraction, allowing further insight into how the extragalactic ultraviolet background (UVB) is created and maintained. In this way, emission line diagnostics simultaneously encode information regarding both galaxy evolution and the drivers of reionization.

We present this foundational work in understanding the conditions of ISM for early galaxies. We introduce the simulations, synthetic spectra, and line diagnostics in Section \ref{sec:ism_methods}. The results of population metallicity and line diagnostics are in Section \ref{sec:ism_results}. Further discussions on the results are in Section \ref{sec:ism_discuss}.

\begin{figure*}[t]
    \centering
    \includegraphics[width=\textwidth]{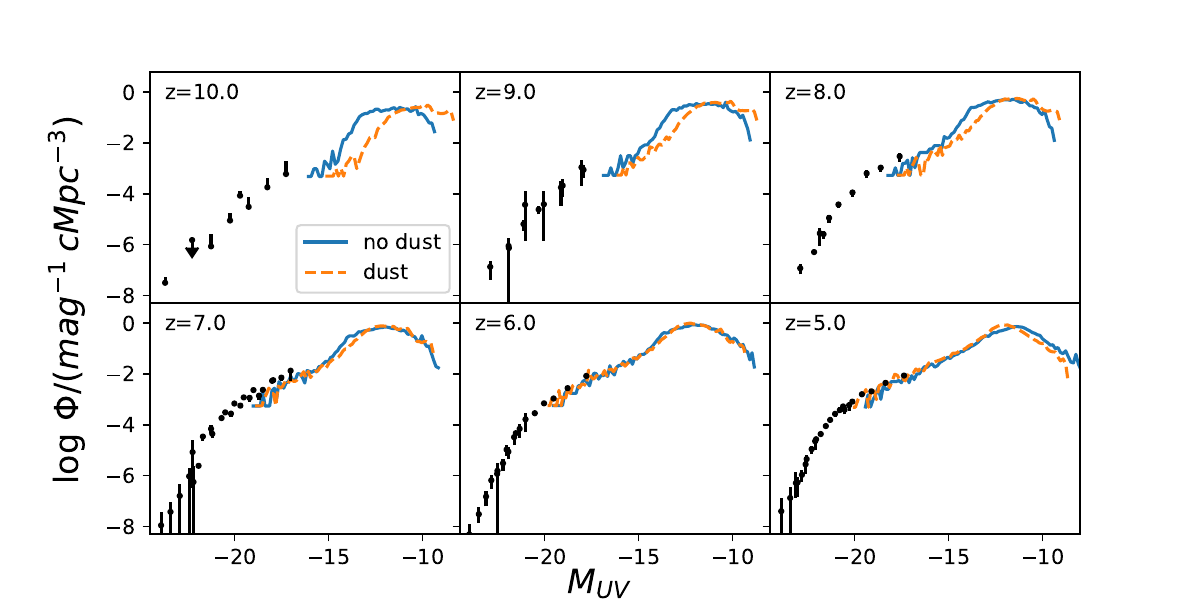}
    \includegraphics[width=\textwidth]{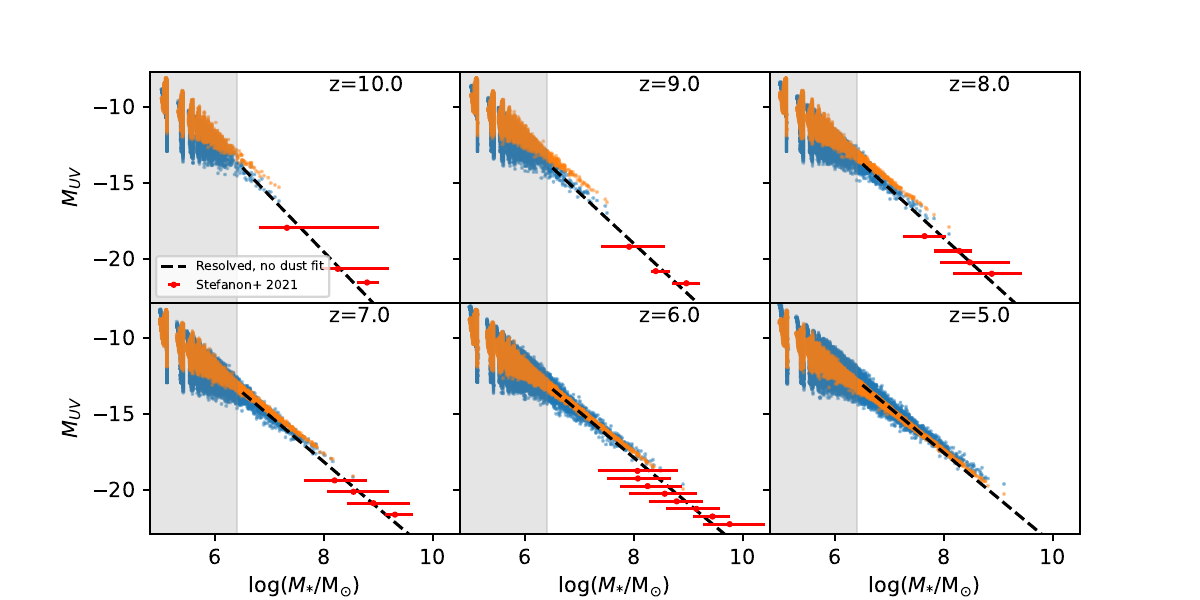}
    \caption{(\textit{Top:}) UV luminosity function calculated from the simulation both with dust (orange dashed) and without dust (blue solid). The scattered points are from \citet{Harikane:2023}. Our simulations probe the low-mass galaxies that have not been observed yet. Agreement is good at $z=5$, but the model underproduces observations by an amount that increases with redshift at $z>5$. (\textit{Bottom:}) Comparison of UV magnitude and stellar mass within our simulation, with estimated fits from the resolved simulated sample (black dashed) alongside observed median bins and 68\% confidence at $z=10$ to $6$ from \citet[][]{Stefanon:2021} after converting the IMF following \citet[][]{Madau:2014}. The gray shaded region is our resolution cut-off of 32 star particles. Simulated galaxies are considered unresolved around magnitudes $M_{\mathrm{UV}} = $ -14 to -13. Extrapolating the simulated mass-to-light ratios yields reasonable agreement with observations.}
    \label{fig:UVLF}
\end{figure*}

\begin{figure}
    \centering
    \includegraphics[width=0.5\textwidth]{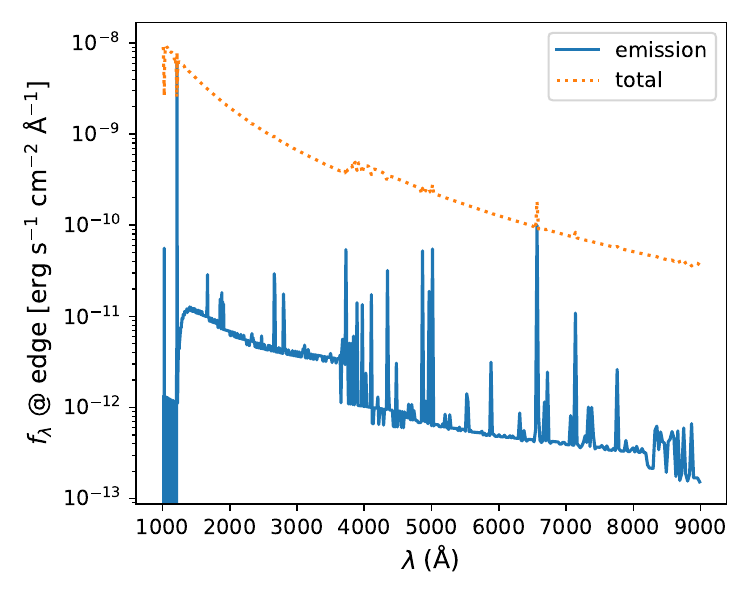}
    \caption{Example spectrum outputted from our {\texttt{Cloudy}} runs. Plotted is the total spectrum (gas and star) and the emission spectrum (gas). The flux density is calculated at the ``edge'' of the galaxy, where the end of the gridding is located.}
    \label{fig:ex_spectrum}
\end{figure}

\section{Methodology} \label{sec:ism_methods}

This section outlines our cosmological simulations, ISM identification, synthetic emission spectra and stellar continua generation, and line diagnostic analysis.

\subsection{\texttt{Technicolor Dawn} Simulations} \label{ssec:method_td}

We used the radiative SPH code \texttt{Technicolor Dawn} \citep[\TD,][]{Finlator:2018} including updates to ionizing escape fraction and feedback \citep{Huscher:2024, Finlator:2020}. It adopts a $\mathrm{\Lambda CDM}$ cosmology with $h = 0.6774$, $\Omega_{\mathrm{\Lambda}} = 0.6911$, $\Omega_m = 0.3089$, $\Omega_b = 0.0486$, and $X_H = 0.751$. The simulation handles gas feedback using a Monte Carlo method based on stellar formation and mass. Stellar information is from the ``Binary Population and Spectral Synthesis'' library, \texttt{BPASS}, v2.2.1 \citep{Stanway:2018} while metal yields (AGB; Type Ia, Type II supernovae or SNe) are from internal tables. Metals are tracked through 10 elements: carbon (C), oxygen (O), silicon (Si), iron (Fe), nitrogen (N), neon (Ne), magnesium (Mg), sulfur (S), calcium (Ca), and titanium (Ti). The metal enrichment uses gas particles' star formation rates for Type II SNe with an assumed hypernovae fraction, $f_\mathrm{HNe}$, and star particle yield rates for Type Ia SNe and AGB stars consistent with previous runs. For a more complete discussion, see~\citet{Finlator:2020}. 


\subsection{Galaxy Synthetic Spectra} \label{ssec:method_spectra}

We identify simulated galaxies with the spline kernel interpolative denmax method \citep[\texttt{SKID},][]{Governato:1997} in star-gas mode and extract gas and stars around the coordinates in our cosmological simulations. A separate intrinsic luminosity calculation was used with \texttt{SKID} outputs to estimate $M_{\mathrm{UV}}$ for Figure \ref{fig:UVLF}  and its method is not based on \texttt{Cloudy} outputs. The gas and stars are reprocessed for each galaxy into $16^3$ fixed cell grids. The size of the grid is determined with a virial-stellar mass relation based on star formation efficiency:

\begin{equation} \label{eqn:mass_relation}
    f_* = \frac{M_*}{M_{\mathrm{vir}}} \frac{\Omega_{\mathrm{M}}}{\Omega_{\mathrm{b}}}
\end{equation}
assuming $f_* = 0.00316$ from the low-mass regime of Model I from \citet[][]{Sun:2016}{}{} since that stays consistent with our simulated dynamic range (Finlator et al., in prep.). We solve for the virial radius and use 5\% of it for the galaxy radius. We smooth gas particles in a Gaussian kernel using each particle's smoothing length onto the grid storing gas density and mass-weighted metallicity. Star particles are checked to see if they reside in the grid space; if so, then we save their coordinates, age, metallicity, and mass. From these grids we perform analyses assuming it is ISM if it meets $n_{\mathrm{H}} \geq 1 $ H atom$/\mathrm{cm}^3$ in the grid cell. 

We run \texttt{Cloudy} v.23.01 \citep[][]{Chatzikos:2023, Ferland:2017}{}{} on the mass resolved galaxy population of $\log M_*/\mathrm{M_\odot} \gtrsim 6.4$ per star particle using the same \texttt{BPASS} model as the simulation to set the incident radiation field. To calculate the incident luminosities, we interpolate the star's isochrone given the age and metallicity from the \texttt{BPASS} tables and the star particle's mass compared to the Kroupa-based \citep[][]{Kroupa:2001} model \texttt{imf.135\_100.} Using the star's coordinates we draw the local cell's gas density and metallicities.

Our decision to run \texttt{Cloudy} in sphere mode obliges us to calibrate the incident flux observationally because our simulations lack the spatial resolution to model the ISM-scale UVB self-consistently. This can be done with line ratios or with the ionization parameter $U$ (or $Q$). Line ratios could impact the metallicity measurements of the simulated spectra, so we ground our calibrations on an averaged observed ionization parameter, $\log Q \sim 8$ [cm/s]. This calibrates our radius for \texttt{Cloudy} runs to 30 ppc each since $Q \propto 1/n_{\mathrm{H}} r^2$ and our ISM densities are similar to other simulated results of around 1--10 cm$^{-3}$ \citep[][]{Katz:2023c}. We run all the parameters on \texttt{Cloudy} for each star particle in the galaxy if it meets the age range $6 \leq \log{t_{*}/yr} \leq 11$. The density is constant throughout the radius of each run and grains are included internally.

With each \texttt{Cloudy }run we cast the incident and total spectrum down a sightline, for this case the simulation's $x$-axis, and attenuate the spectrum within its local cell and cast-through cells. We use the dust-to-mass computations in \citet[][]{Vijayan:2019}{}{} and attenuation computations for $\tau_{\mathrm{dust}}$ and $A_{\mathrm{V}}$ in \citet[][]{Seeyave:2023}{}{} using the model in \citet[][]{Calzetti:2000}{}{} with a fixed $R_V=3.1$. These attenuated spectra are stacked to produce a final spectrum of \species{H}{ii} regions, older stars, and dust -- a composite galactic spectrum.

\begin{figure*}
    \centering
    \includegraphics[width=\textwidth]{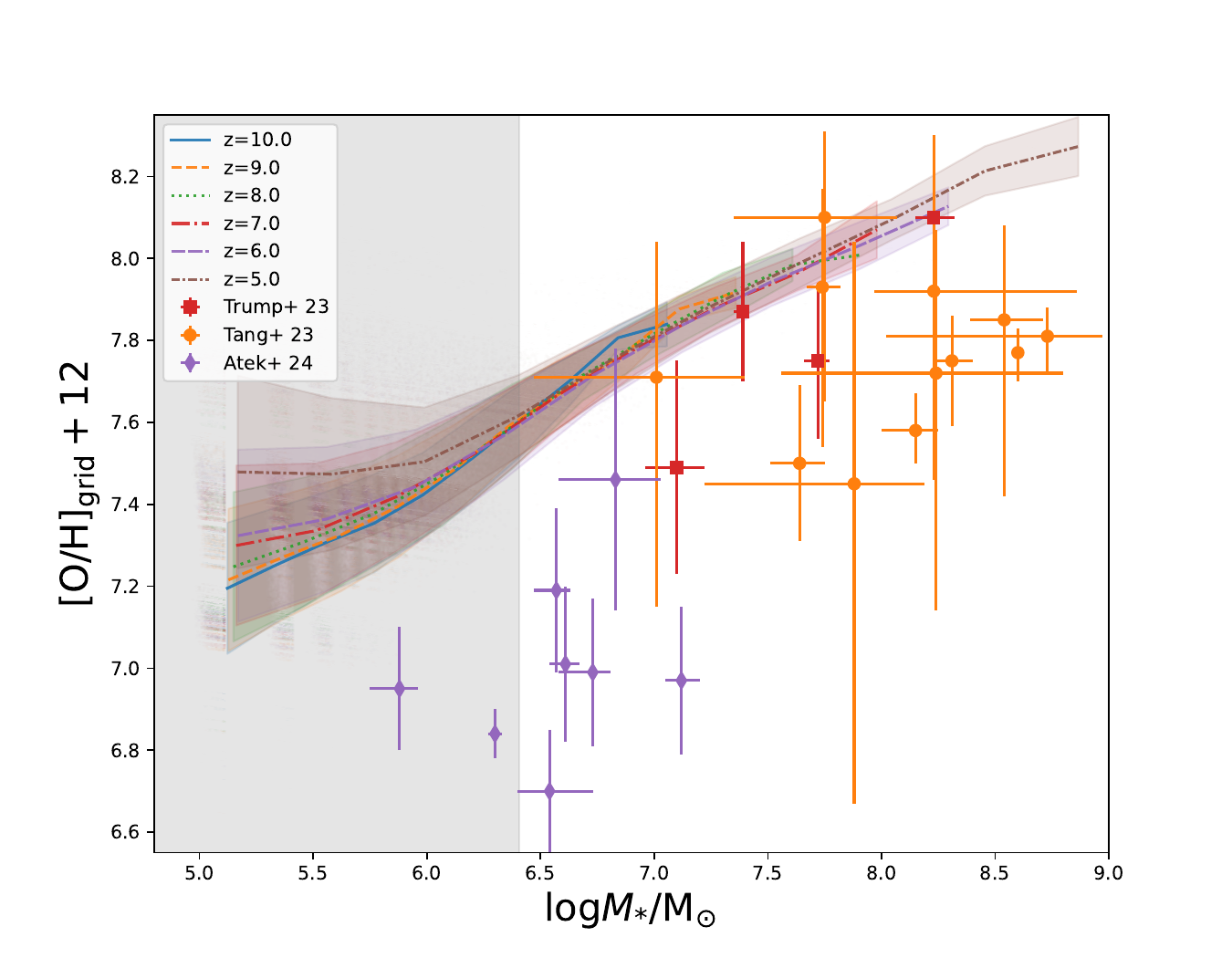}
    \caption{Mass weighted oxygen abundance vs. stellar mass of the galactic population in the simulation. Lines are running median of the population in each redshift with the shaded regions the $1\sigma$ scatter. The oxygen abundance was calculated for each galaxy using the grids, but only counting cells that fit the criteria $n_{\mathrm{H}} \geq 1$ H atom/cm$^{3}$. The gray shaded region denotes data points below the simulation resolution of 32 mean star particles. The scatter points with errorbars are observational data from \citet{Trump:2023} (red squares, $5.27 \lesssim z \lesssim 8.49$), \citet{Tang:2023} (orange circles, $6.92 \lesssim z \lesssim 8.99$), and \citet[][]{Atek:2024} (violet diamonds, $6.88 \lesssim z \lesssim 7.70$) for comparison. Although potentially overlapping considering uncertainties, the simulation tends to a higher bias of oxygen abundance.}
    \label{fig:OH12_mstar_vs_obs}
\end{figure*}

\subsection{Emission Line Ratio Diagnostics} \label{ssec:method_ELDs}
The outputted spectrum includes the total (``net transmitted") and incident spectra using the ``last" option for continuum outputs in \texttt{Cloudy} for each galaxy to get outer radius emission at rest-frame NUV, optical, and NIR. We outputted the rest-frame wavelengths and emission line information as well. We focus on lines of interest from recent observations: $\mathrm{H\beta}$, $\mathrm{H\gamma}$, $[\species{O}{iii}]$5008, $[\species{O}{iii}]4364$, $[\species{O}{ii}]3728$, and $[\species{Ne}{iii}]3870$. We use the flux density value with the wavelength gradient given by \texttt{Cloudy} to calculate line ratios. To get flux from the lines specifically we subtract the incident spectrum's flux density from the total spectrum's flux density to only get emission lines' flux densities. 



\section{Results} \label{sec:ism_results}


\subsection{Galactic population: UVLF and $M_{\mathrm{UV}}-M_*$} \label{ssec:grounding_gal_pop}

In order to justify a detailed comparison between simulated and observed emission lines, we ground-truth our simulated galaxy populations through comparison with the observed rest-frame ultraviolet luminosity function (UVLF) and the stellar mass-luminosity relations. Figure \ref{fig:UVLF} compares the UVLF calculated from stellar luminosities of our star particles with and without dust to observational results compiled in \citet[][]{Harikane:2023}{}{} alongside a mass-luminosity relation plot compared to observed fits in \citet[][]{Stefanon:2021}. Our decision to emphasize relatively high mass resolution leads to weak overlap between the observed and simulated luminosity ranges, but agreement is nonetheless good at $z=$5--6 where direct comparison is possible. A fairly convincing theoretical underproduction arises at $z=7$ and appears to grow with redshift. This may be the low-luminosity manifestation of the well-known early massive galaxy problem \citep[][]{Finkelstein:2024}, which had not yet been identified when this simulation was run. Simulations appear to predict a low-luminosity turnover around $M_{\mathrm{UV}}=-14$ to $-13$ at $z=10$ irrespective of dust, but comparison with the bottom panel confirms that this merely reflects resolution limitations.

The UVLF offset may be due to the mass-luminosity relationship for these galaxies, and they may just be too faint at their given masses. We inspect the $M_*$--$M_{\mathrm{UV}}$ relation to further link our simulated sample to observation. With the fits to the simulated sample, we see similar relations to results from \citet[][]{Stefanon:2021} and this does not seem to be the case that galaxies are too faint. We perform linear fits and calculate slopes of $-0.270 \pm 0.023$ at $z=10$ to $-0.336 \pm 0.002$ at $z=5$ for the resolved sample. The statistical fits are offset from \citet[][]{Stefanon:2021, Santini:2023}, although similar intercepts at $M_{\mathrm{UV}}=-20.5$. However, a different IMF was used between the analyses, which differently weighed the distribution on low-mass stars, so this adds uncertainty with the conversion. Overall, our galaxies are consistent with observed mass-luminosity relations as we are within the median bin uncertainties. 


\subsection{Understanding Link of Metallicity, Oxygen Abundance, and O32} \label{ssec:ism_results_ZOH}


We explored the ISM metallicity with direct outputs through oxygen abundance as it is used extensively as an observational tracer. Figure \ref{fig:OH12_mstar_vs_obs} is the mass-metallicity relation (MZR) using mass-weighted oxygen abundance and the stellar mass of each galaxy within our simulation compared to the current observational results. The simulation predicts no time evolution in [O/H] during the EoR, consistent with \citet[][]{Sanders:2024, Langan:2020}. Observed oxygen abundances are lower than we predict for all masses. 

With this consistent higher bias seen for the simulation, how well is the slope and normalization captured? We performed a $\chi^2$ test between the observations and interpolated fit of the simulation at $z=5$, with low $p$-values indicating that the simulated sample and the observed sample have a similar relation. We get $\log{p} \sim -73$, a strong significance suggesting the simulation captures the slope; however, the simulated result's normalization is higher visually. We repeated the test at $5\sigma$ lower values of the interpolated fit to closer match the observed regime and this nets lower significance at $\log{p} \sim -21$. The simulated results do follow observed results' slope with the MZR, but it is difficult to conclude on the normalization given observed data and visual inferences. Assuming the observations are accurate, the reasonable agreement with the simulated MZR slopes indicates that the way in which feedback efficiency scales with mass is realistic, but that the metal yield is too high or the feedback normalization too low \citep{Finlator:2008}.

\begin{figure*}
    \centering
    \includegraphics[width=0.8\textwidth]{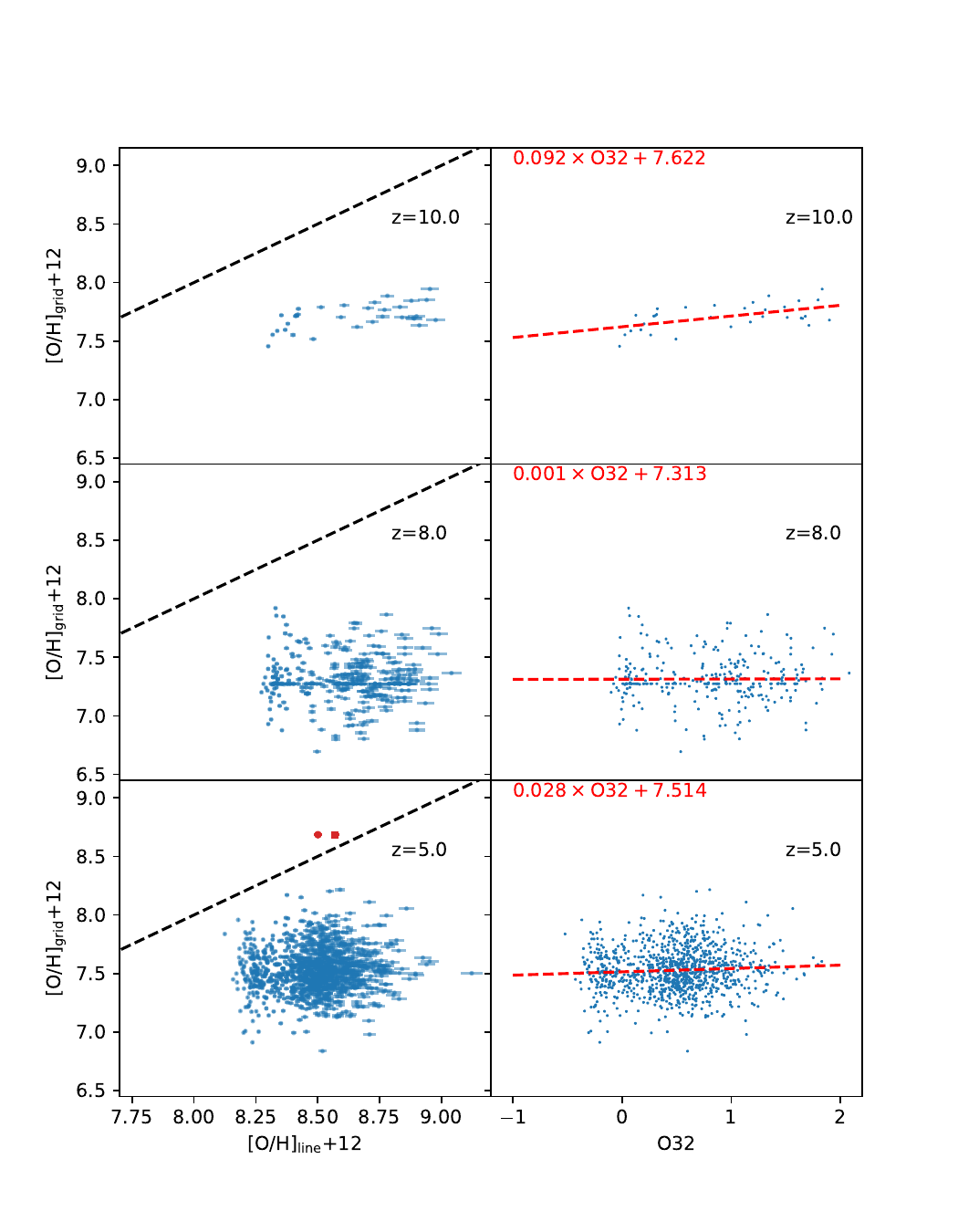}
    \caption{(\textit{Left}) Comparison of oxygen abundance between the grid calculations and \texttt{Cloudy} line ratios using the empirical relation found in \citet{Perez-Montero:2021}. Black, dashed line is a relation of unity. (\textit{Right}) Plot of grid oxygen abundance compared to line ratio O32 from \texttt{Cloudy} spectra with linear fit (red, dashed) alongside equation form (red text). Single \texttt{Cloudy} model outputs are presented with dust-free (red circle) and attentuated with $A_V = 1.1, R_V = 3.1$ (red square) for comparison; calibration errors are at edge of markers. We see large offset of the inferred oxygen abundance, where the empirical calibration is biased higher than our simulation.}
    \label{fig:OH12_gridvcal1}
\end{figure*}

\begin{figure*}
    \centering
    \includegraphics[width=0.9\textwidth]{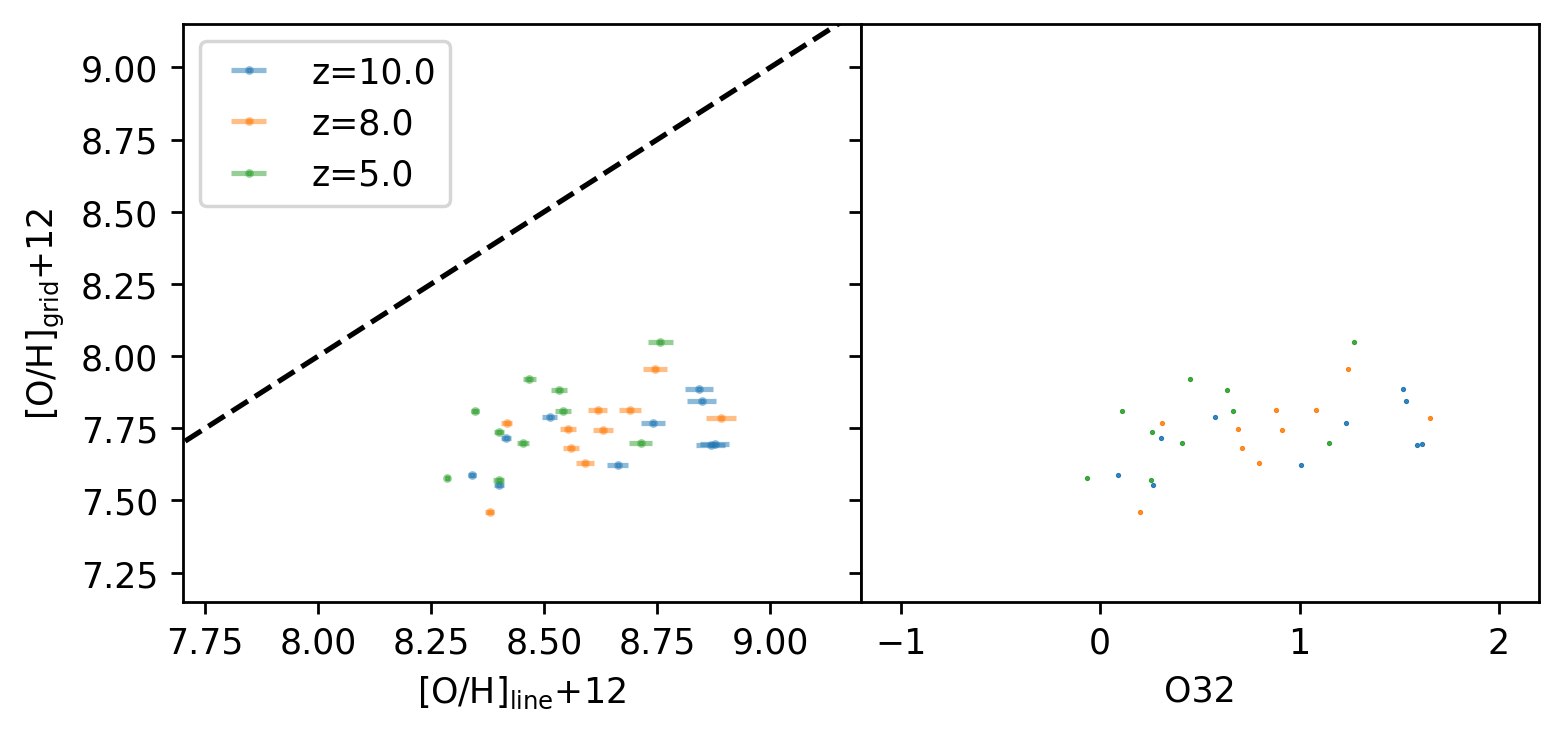}
    \caption{Continuation of Figure \ref{fig:OH12_gridvcal1} but with a subset of galactic spectra without dust attenuation. These dust-free runs also show a significant higher bias in the line calibration estimate.}
    \label{fig:OH12_gridvcal2}
\end{figure*}

An alternative hypothesis for the apparent offset in normalization between simulated and observed MZRs involves systematic uncertainties in strong-line diagnostics. We inspect that concern with the simulation outputs and \texttt{Cloudy} simulated spectra. Figures \ref{fig:OH12_gridvcal1} and \ref{fig:OH12_gridvcal2} compare the oxygen abundance between the simulation outputs and Cloudy line calibration on the left, with O32 \citep[][]{Perez-Montero:2021}{}{} against the simulation outputs on the right. The metallicities inferred from strong-line indicators (abscissa) are biased higher by about 1 dex compared to the mass-weighted, intrinsic values (ordinate). Comparing linearly-regressed fits between the inferred calibration and simulation we find consistent offsets: the simulation has a considerably lower slope, lower normalization, and both show no relations between redshifts. This also coincides with the large scatter between the simulated oxygen abundance and O32, suggesting the line ratio itself is not a strong indicator. Concluding from Figure~\ref{fig:OH12_gridvcal1} that the observationally-inferred metallicities are biased high further exacerbates the difference in the MZR for our simulation and current observation. We note that a majority of our galaxies do match the density regime in which the observational calibration is expected to work best.

One possible explanation for the mismatch between intrinsic and inferred metallicity is that the luminosity-and mass-weighted metallicities differ in a way that does not emerge from single-zone models. We confirm that  single \texttt{Cloudy} model results do not have this issue (see the red circle and square in the bottom-left panel of Figure~\ref{fig:OH12_gridvcal1}). In fact, the single model's strong line calibration slightly under-predicts the intrinsic oxygen abundance. Nor can the offset be attributed entirely to dust: a subset of dust-free runs were performed to compute [O/H] as seen in Figure \ref{fig:OH12_gridvcal2}, and the line calibration still overpredicts [O/H]. It may not be that the relation is inherently wrong, but could point out fundamental differences that EoR galaxies have compared to lower redshift  populations, as well as the added complexity from contaminating starlight in less active regions of the ISM. This also could be an effect from incorrectly accounting for dust attenuation, as dust could impact [\species{O}{ii}] 3728 line intensity more than the [\species{O}{iii}] 4958, 5008 line intensities and boost the ratio.%


\begin{figure*}
    \centering
    \includegraphics{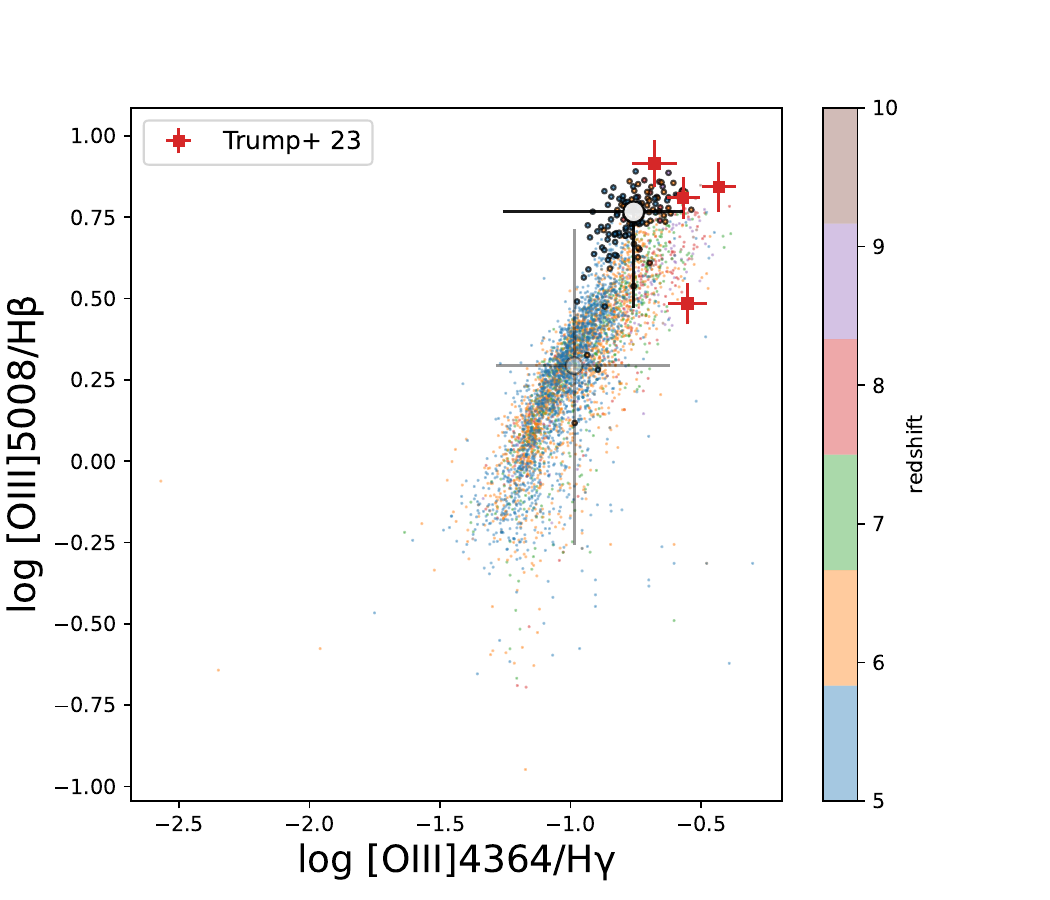}
    \caption{Line diagnostic plot of [OIII]/H$\beta$ vs. [OIII]/H$\gamma$ compared to observed data in \citet[][]{Trump:2023}{}{}. Simulated sample (circles) are split between those $M_{\mathrm{UV}} < -16.9$ (larger, black outline, more opaque circles) than the rest of the simulated sample (smaller, no outline, more transparent circles) to show potentially observable galaxies given JWST limits \cite{Tang:2023}. Two median points of the simulated sample with 95\% confidence are plotted as white circles: for the whole sample (more transparent, smaller) and the $M_{\mathrm{UV}} < -16.9$ sample (larger, more opaque). Some simulated points lie within errors of observed data and especially in potentially observable ranges, but not most, where most [\species{O}{iii}] lines are weaker than their respective Balmer line.}
    \label{fig:O3Hb_O3Hg}
\end{figure*}

\begin{figure*}
    \centering
    \includegraphics{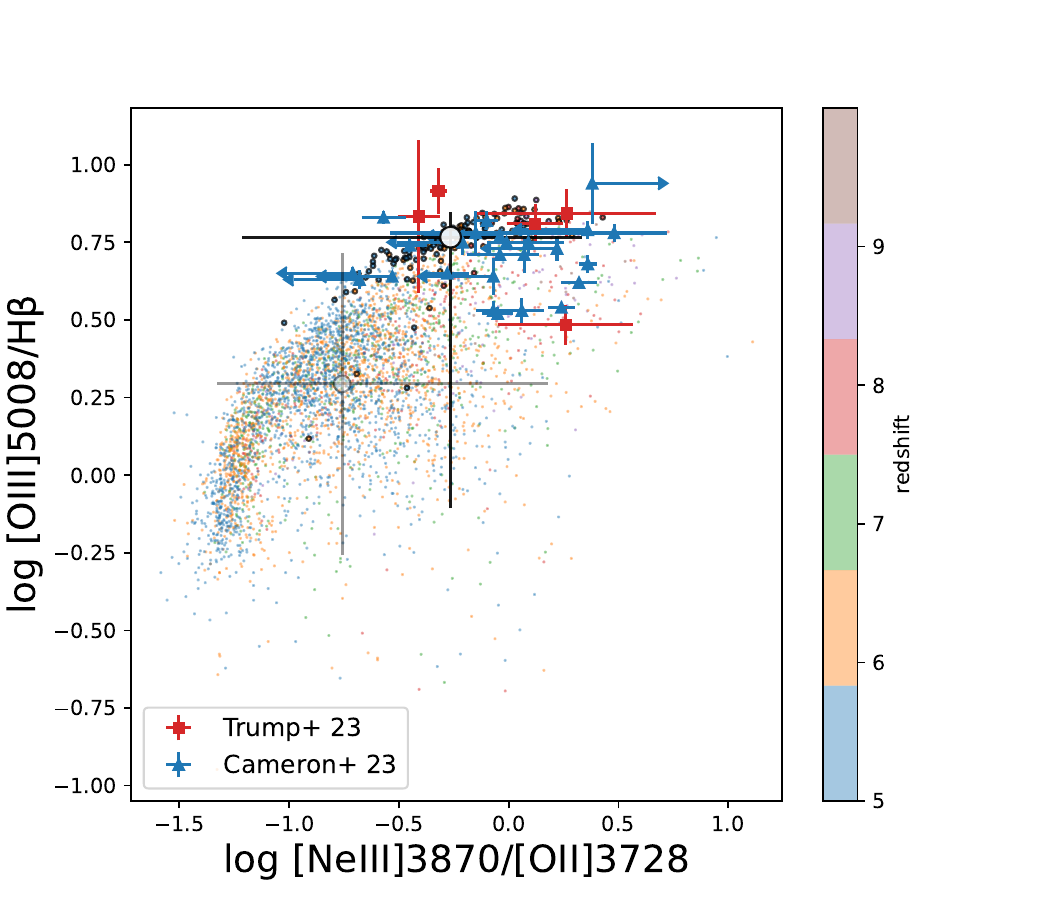}
    \caption{Line diagnostic plot of [OIII]5008/H$\beta$ (O3) vs. [NeIII]3870/[OII]3728 (Ne3O2) compared to observed data in \citet{Trump:2023} (red squares) and \citet{Cameron:2023} (blue triangles). Simulated sample (circles) are split between those $M_{\mathrm{UV}} < -16.9$ (larger, black outline, more opaque circles) than the rest of the simulated sample (smaller, no outline, more transparent circles) to show potentially observable galaxies given JWST limits \cite{Tang:2023}. Two median points of the simulated sample with 95\% confidence are plotted as white circles: for the whole sample (more transparent, smaller) and the $M_{\mathrm{UV}} < -16.9$ sample (larger, more opaque). The simulated and observed galaxies have similar ranges of Ne3O2, but still simulated galaxies have weaker O3 compared to the whole sample, but there is considerable overlap with potentially observable galaxies.}
    \label{fig:O3_Ne3O2}
\end{figure*}

\subsection{Emission Line Diagnostics} \label{ssec:ELD_results}


The \texttt{Cloudy} runs are calibrated such that the mode of the ionization parameters matches observations. What is expected for the galactic population as a whole? We need to investigate emission line diagnostics. Continuing the focus on oxygen, we explore the ratios of [\species{O}{iii}]5008 and 4364 $\mathrm{\AA}$ to \species{H}{$\beta$}  and \species{H}{$\gamma$} respectively. Figure \ref{fig:O3Hb_O3Hg} is the diagnostic plot of [\species{O}{iii}]5007/\species{H}{$\beta$} to [\species{O}{iii}]4364/\species{H}{$\gamma$} against recent observational results. Darker and lighter circles denote simulated galaxies that are brighter and fainter than the observational limit of $M_{UV}<-16.9$~\citep{Tang:2023}, respectively. First, within the simulated data there appears to be no evolution in line ratios with redshift, meaning no evidence of enrichment level over time. This is consistent with Figure \ref{fig:OH12_mstar_vs_obs}. Second, while the line ratios for simulated bright galaxies (heavier circles) fall much closer to observations than fainter galaxies do, they are still biased to low [OIII]/H$\gamma$ and to a lesser extent in [OIII]/H$\beta$. 

This inconsistency seems contradictory to our simulation's high oxygen abundances. Are the local ionization fields playing a prominent role? Figure \ref{fig:O3_Ne3O2} plots [\species{O}{iii}]5007/\species{H}{$\beta$} vs. [\species{Ne}{iii}]3870/[\species{O}{ii}]3728 (O3 vs. Ne3O2). Similarly here first, we do not see an evolution of Ne3O2 to redshift like the results from Figure \ref{fig:OH12_mstar_vs_obs} suggest. More importantly, the Ne3O2 seem similar for both simulated and observed galactic samples including in observable range with similar O3 for bright galaxies. However, the overall simulated sample does have weaker O3 points.

Overall, the simulated line diagnostics suggest a population of lower ionization, but that differs from the direct outputs from \texttt{Cloudy}. Figure \ref{fig:q_dist} shows the distribution of each galaxy's average ionization parameter $Q$ at each redshift. Although somewhat consistent in one of the modes being around our hypothesis, our models also allow a highly ionized population of galaxies with $\log Q \geq 9$ [cm/s]. Based on our derived distribution, our galaxies exhibit higher ionization.

\section{Discussion} \label{sec:ism_discuss}

\subsection{Metal Enrichment in ISM} \label{ssec:ism_discuss_Z_ISM}

Our models produce more oxygen overall compared to present observations.  Although statistically significant, visual evidence of simulated results shows predictions occupying around the upper limits of observational errors. To match MZR normalizations we either reduce the SNe rates, reduce the expected $f_{\mathrm{HNe}}$, reduce SFR, or reduce metal yields. Reducing SFR reduces the population's stellar mass and therefore the normalization of the UVLF, which we are currently consistent with. Reducing SNe rates or $f_{\mathrm{HNe}}$ reduces outflows from our simulations and exacerbate the ``carbon problem" seen in our metal absorbers while introducing tension with the observed abundance of circumgalactic silicon, which the simulation currently repduces reasonably well~\citep[][]{Huscher:2024, D'Odorico:2022, Finlator:2020}. As such, we predict this is due to a difference in oxygen yields within our simulations compared to observations. Otherwise, the offset may be with observations given the direct calibrations can still underpredict [O/H] given only rest-frame emissions \citep[][]{Harikane:2025}. The simulated oxygen yield could be too high because the assumed initial mass function (IMF) is incorrect, and this has testable consequences. If the the correct IMF is weighted more heavily to massive stars, then the correct oxygen yield is lower while the silicon yield is higher, with implications for the O/Si ratio as well as predicted ionization parameters~\citep{Kulkarni:2013}. Likewise, if the correct IMF is weighted to lower-mass stars, then the (C/O) yields (increase/decrease), with associated implications for associated emission and absorption features.

Another result is the non-evolving MZR normalization lacking evolution of metallicity in fixed stellar mass. This lack of evolution has been predicted previously~\citep{Langan:2020} and is reflected in observations \citep[][]{Nakajima:2023}{}{}. It indicates a non-evolving balance between enrichment and dilution, although a conspiracy between, for example, simultaneously increasing (or decreasing) metal yield and feedback efficiency is not ruled out.

The use of O32 to trace oxygen abundance in observation is in conflict with our simulation results alongside assumed ionization within the galactic population. The observational calibration is based on \citet[][]{Perez-Montero:2021} and the offset is seen in previous works and show that the O32 calibration is dependent on $Q$ \citep[][]{Hirschmann:2023}. Although it may seem questionable to compare with cosmological simulations due to numerical offsets or resolution limitations, our densities and temperatures are within the limits proposed in the calibration. The single-zone models are more in agreement with such calibrations. This puts into question the validity of using low-redshift calibrations in the high-redshift regime if the models cannot come to an agreement with observation, especially considering the various ionization fields.

\subsection{Line Diagnostics Interpretation}

\begin{figure}
    \centering
    \includegraphics[width=0.99\linewidth]{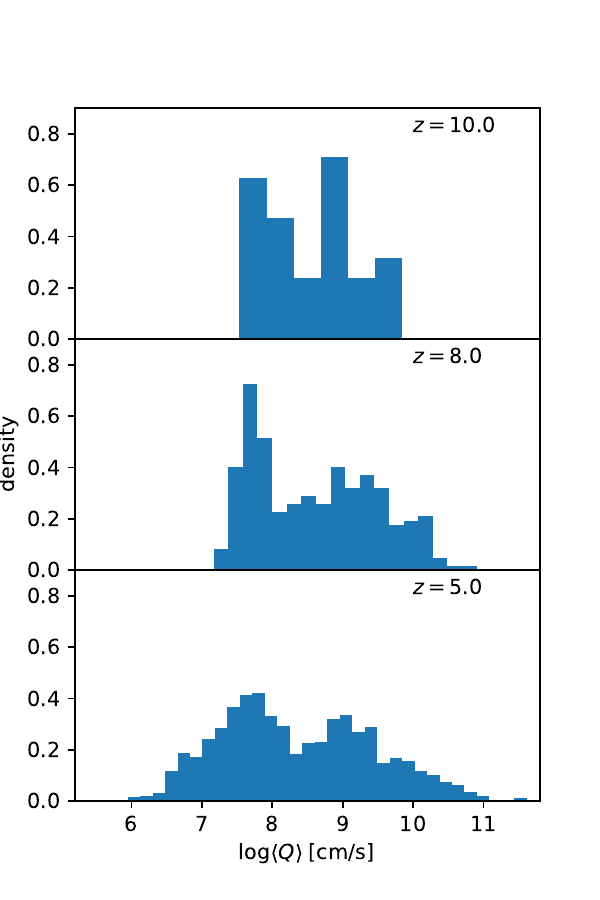}
    \caption{Distribution of mean ionization parameter $Q$ for various redshifts. We see that we stay consistent with a modal $\log(Q/\mathrm{cm\,s^{-1})} \sim 8$, but higher ionization galaxies exist within our assumption.}
    \label{fig:q_dist}
\end{figure}

With [O/H] being higher, why are simulated galaxies producing relatively weaker [\species{O}{iii}] lines? Prior predictions \citep[][]{Trump:2023} suggest lower ionization and higher pressure, but this also likely stems from higher ionization background of these \species{H}{ii} regions as seen with Figure \ref{fig:q_dist}. Our models start with constraining $Q$ for the population's mode to the observational results; this leaves some galaxies to become highly ionized to $\log Q \sim 11$. These regions likely ionize up to \species{O}{iv} and higher. As such, we expect there to be harsher ionizing flux produced from the galactic population. This now allows testing of our model \species{H}{ii} regions if future observations stay consistent and disprove us. Additionally, a search into high-ionization lines can help break the contradiction, for example \species{Ne}{v} or \species{He}{ii} lines.

However, the observational data points do overlap with our simulated sample when only inferring from the simulated galaxies that can be observed. As such, the population's ionization inferred from the simulations could still be true for observations given their specific magnitude range. If so, many of the dwarf galaxies within the galactic population have ionizing parameters far higher at $\log Q \sim 9$ [cm/s] indicating higher ionizing photon production from them. However, further work must be done as such population comparisons likely have degenerate answers where different distributions can produce the same line diagnostic results.

\subsection{Caveats} \label{ssec:ism_discuss_caveats}

One concern is how we constrained the ionization parameter of our \texttt{Cloudy} models. Radius was chosen as our number densities are comparable to other reported number densities of around 1 to 100 cm$^{-1}$ \citep[][]{Katz:2023c}{}{}. The calibration is based on a physical parameter, $Q$, to match inferred measurements thus far and assuming they are the most abundant galaxies, i.e. the mode of the galactic population. This is the initial comparison for the model's null hypothesis. However, this does not include unobserved galaxies that may be too faint to see. As such, we need to further test the impact on reionization if most galaxies have harsher ionizing fields in their ISM.

Another concern is how we constrained with $Q$ instead of line ratios. Constraining $Q$ is similar to constraining Ne3O2 and [\species{O}{iii}]/\species{H}{$\beta$} since observational calibrations of the ionization parameter used those lines. It should not be surprising that we overlap with Ne3O2 and O3, but we have two issues: higher metallicity with lower [\species{O}{iii}] in Figure \ref{fig:O3Hb_O3Hg}, and the discrepancy with O32 in Figure \ref{fig:OH12_gridvcal1}. Most galaxies would likely be lower ionized, higher-pressure systems, but the distribution of $Q$ in Figure \ref{fig:q_dist} suggests otherwise. The discrepancy with O32 is concerning, as this is partially calibrated by [\species{O}{iii}]5008 and [\species{O}{ii}]3728, which may mean stronger ionization within \species{H}{ii} regions, or potentially other unknown processes being prevalent.

The gridding done on each galaxy is larger than the resolution of the \species{H}{ii} regions simulated in \texttt{Cloudy}. These runs had to be calibrated to be consistent with observed and simulated values for the ionization parameter and densities, as has been stated. We do not believe that the runs have been compromised. However, this is a limitation that can be tackled in future work with either zoom-in simulations or more powerful tools in cosmological simulations.

On the observational end, there may be contamination from the AGN population, which should impact emission lines observed. Aside from the little red dots (LRD), some galaxies may host AGN without immediate evidence \citep{Davis:2024}. To account for AGN there needs to be follow-up data with either confirmation of lines and/or non-stellar rest-frame UV continuum or photometric calibration of point sources. 

\section{Conclusion} \label{sec:conc}

We present our results on simulation outputs and post-processing synthetic spectra modeling using \texttt{Cloudy}. We find emission line diagnostics for early galaxies at $10 \geq z \geq 5$ near the end of Reionization and study the metallicity and ionization within \species{H}{ii} regions of these galaxies. We find the following:

\begin{enumerate}
    \item Simulation results predict higher oxygen abundances compared to what was inferred. The slopes seem consistent, so feedback mechanisms seem to be captured well. However, yields are likely predicted higher given the normalization.
    \item There is a large systematic difference between simulation estimates and strong line calibration estimates for oxygen abundance. This is reflected in scatter of O32 and underestimations of single-model runs compared to overestimation with the composite spectra. This suggests a complicated issue discerning starlight around less active regions of the ISM and correct dust modeling.
    \item Our simulated spectra have weaker [\species{O}{iii}] lines but comparable Ne3O2. Given the assumption that our mode of the galactic population can be matching observation, this implies the existence of galaxies with higher $Q$. This suggests the population is producing harsher ionizing galaxies than observed.
\end{enumerate}

\begin{acknowledgments}
S. Ku\v{s}mi\'{c} is supported by the National Science Foundation (NSF) under Award Number 2006550. The Technicolor Dawn simulations were enabled by the Extreme Science and Engineering Discovery Environment (XSEDE), which is supported by NSF grant number ACI-1548562, now transferred services to Advanced Cyberinfrastructure Coordination Ecosystem: Services \& Support (ACCESS). The Cosmic Dawn Center is funded by the Danish National Research Foundation. Generative AI \texttt{Chat-GPT 4o} has been used only for code development with new modules, such as \texttt{concurrent}.

We would like to thank Allesandra Venditti for helping understand \texttt{Cloudy} predictions during conference talks. S. Ku\v{s}mi\'{c} contributed with the data analysis coding, experimental design, post-processing with \texttt{Cloudy}, and primary drafting. K. Finlator contributed by providing feedback to the manuscript, processing the \texttt{TD} runs and outputs, and guidance in the experiment. E. Huscher contributed by providing feedback to the manuscript. M. Steen contributed by providing feedback to the manuscript.
\end{acknowledgments}

\software{\texttt{numpy} \citep{Harris:2020},  
          \texttt{Cloudy} \citep{Chatzikos:2023}, 
          \texttt{scipy} \citep{Virtanen:2020},
          \texttt{yt} \citep{Turk:2011},
          \texttt{pycosie} \citep{kusmic_2024_pycosie}
          }

\bibliography{ref}{}

\begin{thebibliography}{}
\expandafter\ifx\csname natexlab\endcsname\relax\def\natexlab#1{#1}\fi
\providecommand{\url}[1]{\href{#1}{#1}}
\providecommand{\dodoi}[1]{doi:~\href{http://doi.org/#1}{\nolinkurl{#1}}}
\providecommand{\doeprint}[1]{\href{http://ascl.net/#1}{\nolinkurl{http://ascl.net/#1}}}
\providecommand{\doarXiv}[1]{\href{https://arxiv.org/abs/#1}{\nolinkurl{https://arxiv.org/abs/#1}}}

\bibitem[{{Atek} {et~al.}(2024){Atek}, {Labb{\'e}}, {Furtak}, {Chemerynska}, {Fujimoto}, {Setton}, {Miller}, {Oesch}, {Bezanson}, {Price}, {Dayal}, {Zitrin}, {Kokorev}, {Weaver}, {Brammer}, {Dokkum}, {Williams}, {Cutler}, {Feldmann}, {Fudamoto}, {Greene}, {Leja}, {Maseda}, {Muzzin}, {Pan}, {Papovich}, {Nelson}, {Nanayakkara}, {Stark}, {Stefanon}, {Suess}, {Wang}, \& {Whitaker}}]{Atek:2024}
{Atek}, H., {Labb{\'e}}, I., {Furtak}, L.~J., {et~al.} 2024, \nat, 626, 975, \dodoi{10.1038/s41586-024-07043-6}

\bibitem[{{Calzetti} {et~al.}(2000){Calzetti}, {Armus}, {Bohlin}, {Kinney}, {Koornneef}, \& {Storchi-Bergmann}}]{Calzetti:2000}
{Calzetti}, D., {Armus}, L., {Bohlin}, R.~C., {et~al.} 2000, \apj, 533, 682, \dodoi{10.1086/308692}

\bibitem[{{Cameron} {et~al.}(2023){Cameron}, {Saxena}, {Bunker}, {D'Eugenio}, {Carniani}, {Maiolino}, {Curtis-Lake}, {Ferruit}, {Jakobsen}, {Arribas}, {Bonaventura}, {Charlot}, {Chevallard}, {Curti}, {Looser}, {Maseda}, {Rawle}, {Rodr{\'\i}guez Del Pino}, {Smit}, {{\"U}bler}, {Willott}, {Witstok}, {Egami}, {Eisenstein}, {Johnson}, {Hainline}, {Rieke}, {Robertson}, {Stark}, {Tacchella}, {Williams}, {Willmer}, {Bhatawdekar}, {Bowler}, {Boyett}, {Circosta}, {Helton}, {Jones}, {Kumari}, {Ji}, {Nelson}, {Parlanti}, {Sandles}, {Scholtz}, \& {Sun}}]{Cameron:2023}
{Cameron}, A.~J., {Saxena}, A., {Bunker}, A.~J., {et~al.} 2023, \aap, 677, A115, \dodoi{10.1051/0004-6361/202346107}

\bibitem[{{Chatzikos} {et~al.}(2023){Chatzikos}, {Bianchi}, {Camilloni}, {Chakraborty}, {Gunasekera}, {Guzm{\'a}n}, {Milby}, {Sarkar}, {Shaw}, {van Hoof}, \& {Ferland}}]{Chatzikos:2023}
{Chatzikos}, M., {Bianchi}, S., {Camilloni}, F., {et~al.} 2023, \rmxaa, 59, 327, \dodoi{10.22201/ia.01851101p.2023.59.02.12}

\bibitem[{{Davis} {et~al.}(2024){Davis}, {Trump}, {Simons}, {McGrath}, {Wilkins}, {Arrabal Haro}, {Bagley}, {Dickinson}, {Fern{\'a}ndez}, {Amor{\'\i}n}, {Backhaus}, {Cleri}, {Llerena}, {Brunker}, {Barro}, {Bisigello}, {Brooks}, {Costantin}, {de La Vega}, {Dekel}, {Finkelstein}, {Hathi}, {Hirschmann}, {Kartaltepe}, {Koekemoer}, {Lucas}, {Papovich}, {P{\'e}rez-Gonz{\'a}lez}, {Pirzkal}, {Rodighiero}, {Rose}, {Yung}, \& {Ceers Collaborators}}]{Davis:2024}
{Davis}, K., {Trump}, J.~R., {Simons}, R.~C., {et~al.} 2024, \apj, 974, 42, \dodoi{10.3847/1538-4357/ad6865}

\bibitem[{{D'Odorico} {et~al.}(2022){D'Odorico}, {Finlator}, {Cristiani}, {Cupani}, {Perrotta}, {Calura}, {C{\`e}nturion}, {Becker}, {Berg}, {Lopez}, {Ellison}, \& {Pomante}}]{D'Odorico:2022}
{D'Odorico}, V., {Finlator}, K., {Cristiani}, S., {et~al.} 2022, \mnras, 512, 2389, \dodoi{10.1093/mnras/stac545}

\bibitem[{{Ferland} {et~al.}(2017){Ferland}, {Chatzikos}, {Guzm{\'a}n}, {Lykins}, {van Hoof}, {Williams}, {Abel}, {Badnell}, {Keenan}, {Porter}, \& {Stancil}}]{Ferland:2017}
{Ferland}, G.~J., {Chatzikos}, M., {Guzm{\'a}n}, F., {et~al.} 2017, \rmxaa, 53, 385, \dodoi{10.48550/arXiv.1705.10877}

\bibitem[{{Finkelstein} {et~al.}(2023){Finkelstein}, {Bagley}, {Ferguson}, {Wilkins}, {Kartaltepe}, {Papovich}, {Yung}, {Arrabal Haro}, {Behroozi}, {Dickinson}, {Kocevski}, {Koekemoer}, {Larson}, {Le Bail}, {Morales}, {P{\'e}rez-Gonz{\'a}lez}, {Burgarella}, {Dav{\'e}}, {Hirschmann}, {Somerville}, {Wuyts}, {Bromm}, {Casey}, {Fontana}, {Fujimoto}, {Gardner}, {Giavalisco}, {Grazian}, {Grogin}, {Hathi}, {Hutchison}, {Jha}, {Jogee}, {Kewley}, {Kirkpatrick}, {Long}, {Lotz}, {Pentericci}, {Pierel}, {Pirzkal}, {Ravindranath}, {Ryan}, {Trump}, {Yang}, {Bhatawdekar}, {Bisigello}, {Buat}, {Calabr{\`o}}, {Castellano}, {Cleri}, {Cooper}, {Croton}, {Daddi}, {Dekel}, {Elbaz}, {Franco}, {Gawiser}, {Holwerda}, {Huertas-Company}, {Jaskot}, {Leung}, {Lucas}, {Mobasher}, {Pandya}, {Tacchella}, {Weiner}, \& {Zavala}}]{Finkelstein:2023}
{Finkelstein}, S.~L., {Bagley}, M.~B., {Ferguson}, H.~C., {et~al.} 2023, \apjl, 946, L13, \dodoi{10.3847/2041-8213/acade4}

\bibitem[{{Finkelstein} {et~al.}(2024){Finkelstein}, {Leung}, {Bagley}, {Dickinson}, {Ferguson}, {Papovich}, {Akins}, {Arrabal Haro}, {Dav{\'e}}, {Dekel}, {Kartaltepe}, {Kocevski}, {Koekemoer}, {Pirzkal}, {Somerville}, {Yung}, {Amor{\'\i}n}, {Backhaus}, {Behroozi}, {Bisigello}, {Bromm}, {Casey}, {Ch{\'a}vez Ortiz}, {Cheng}, {Chworowsky}, {Cleri}, {Cooper}, {Davis}, {de la Vega}, {Elbaz}, {Franco}, {Fontana}, {Fujimoto}, {Giavalisco}, {Grogin}, {Holwerda}, {Huertas-Company}, {Hirschmann}, {Iyer}, {Jogee}, {Jung}, {Larson}, {Lucas}, {Mobasher}, {Morales}, {Morley}, {Mukherjee}, {P{\'e}rez-Gonz{\'a}lez}, {Ravindranath}, {Rodighiero}, {Rowland}, {Tacchella}, {Taylor}, {Trump}, \& {Wilkins}}]{Finkelstein:2024}
{Finkelstein}, S.~L., {Leung}, G. C.~K., {Bagley}, M.~B., {et~al.} 2024, \apjl, 969, L2, \dodoi{10.3847/2041-8213/ad4495}

\bibitem[{{Finlator} \& {Dav{\'e}}(2008)}]{Finlator:2008}
{Finlator}, K., \& {Dav{\'e}}, R. 2008, \mnras, 385, 2181, \dodoi{10.1111/j.1365-2966.2008.12991.x}

\bibitem[{{Finlator} {et~al.}(2020){Finlator}, {Doughty}, {Cai}, \& {D{\'\i}az}}]{Finlator:2020}
{Finlator}, K., {Doughty}, C., {Cai}, Z., \& {D{\'\i}az}, G. 2020, \mnras, 493, 3223, \dodoi{10.1093/mnras/staa377}

\bibitem[{{Finlator} {et~al.}(2018){Finlator}, {Keating}, {Oppenheimer}, {Dav{\'e}}, \& {Zackrisson}}]{Finlator:2018}
{Finlator}, K., {Keating}, L., {Oppenheimer}, B.~D., {Dav{\'e}}, R., \& {Zackrisson}, E. 2018, \mnras, 480, 2628, \dodoi{10.1093/mnras/sty1949}

\bibitem[{{Governato} {et~al.}(1997){Governato}, {Moore}, {Cen}, {Stadel}, {Lake}, \& {Quinn}}]{Governato:1997}
{Governato}, F., {Moore}, B., {Cen}, R., {et~al.} 1997, \na, 2, 91, \dodoi{10.1016/S1384-1076(97)00011-0}

\bibitem[{{Harikane} {et~al.}(2023){Harikane}, {Ouchi}, {Oguri}, {Ono}, {Nakajima}, {Isobe}, {Umeda}, {Mawatari}, \& {Zhang}}]{Harikane:2023}
{Harikane}, Y., {Ouchi}, M., {Oguri}, M., {et~al.} 2023, \apjs, 265, 5, \dodoi{10.3847/1538-4365/acaaa9}

\bibitem[{{Harikane} {et~al.}(2025){Harikane}, {Sanders}, {Ellis}, {Jones}, {Ouchi}, {Laporte}, {Roberts-Borsani}, {Katz}, {Nakajima}, {Ono}, \& {Gupta}}]{Harikane:2025}
{Harikane}, Y., {Sanders}, R.~L., {Ellis}, R., {et~al.} 2025, arXiv e-prints, arXiv:2505.09186, \dodoi{10.48550/arXiv.2505.09186}

\bibitem[{{Harris} {et~al.}(2020){Harris}, {Millman}, {van der Walt}, {Gommers}, {Virtanen}, {Cournapeau}, {Wieser}, {Taylor}, {Berg}, {Smith}, {Kern}, {Picus}, {Hoyer}, {van Kerkwijk}, {Brett}, {Haldane}, {del R{\'\i}o}, {Wiebe}, {Peterson}, {G{\'e}rard-Marchant}, {Sheppard}, {Reddy}, {Weckesser}, {Abbasi}, {Gohlke}, \& {Oliphant}}]{Harris:2020}
{Harris}, C.~R., {Millman}, K.~J., {van der Walt}, S.~J., {et~al.} 2020, \nat, 585, 357, \dodoi{10.1038/s41586-020-2649-2}

\bibitem[{{Hirschmann} {et~al.}(2023){Hirschmann}, {Charlot}, \& {Somerville}}]{Hirschmann:2023}
{Hirschmann}, M., {Charlot}, S., \& {Somerville}, R.~S. 2023, \mnras, 526, 3504, \dodoi{10.1093/mnras/stad2745}

\bibitem[{{Hu} {et~al.}(2024){Hu}, {Papovich}, {Dickinson}, {Kennicutt}, {Shen}, {Amor{\'\i}n}, {Arrabal Haro}, {Bagley}, {Bhatawdekar}, {Cleri}, {Cole}, {Dekel}, {de la Vega}, {Finkelstein}, {Grogin}, {Hathi}, {Hirschmann}, {Holwerda}, {Hutchison}, {Jung}, {Koekemoer}, {Kartaltepe}, {Lucas}, {Llerena}, {Mascia}, {Mobasher}, {Napolitano}, {Newman}, {Pentericci}, {P{\'e}rez-Gonz{\'a}lez}, {Trump}, {Wilkins}, \& {Yung}}]{Hu:2024}
{Hu}, W., {Papovich}, C., {Dickinson}, M., {et~al.} 2024, \apj, 971, 21, \dodoi{10.3847/1538-4357/ad5015}

\bibitem[{{Huscher} {et~al.}(2024){Huscher}, {Finlator}, {Ku{\v{s}}mi{\'c}}, \& {Steen}}]{Huscher:2024}
{Huscher}, E., {Finlator}, K., {Ku{\v{s}}mi{\'c}}, S., \& {Steen}, M. 2024, arXiv e-prints, arXiv:2404.00193, \dodoi{10.48550/arXiv.2404.00193}

\bibitem[{{Katz} {et~al.}(2023){Katz}, {Saxena}, {Rosdahl}, {Kimm}, {Blaizot}, {Garel}, {Michel-Dansac}, {Haehnelt}, {Ellis}, {Pentericci}, {Devriendt}, \& {Slyz}}]{Katz:2023c}
{Katz}, H., {Saxena}, A., {Rosdahl}, J., {et~al.} 2023, \mnras, 518, 270, \dodoi{10.1093/mnras/stac3019}

\bibitem[{{Kroupa}(2001)}]{Kroupa:2001}
{Kroupa}, P. 2001, \mnras, 322, 231, \dodoi{10.1046/j.1365-8711.2001.04022.x}

\bibitem[{{Kulkarni} {et~al.}(2013){Kulkarni}, {Rollinde}, {Hennawi}, \& {Vangioni}}]{Kulkarni:2013}
{Kulkarni}, G., {Rollinde}, E., {Hennawi}, J.~F., \& {Vangioni}, E. 2013, \apj, 772, 93, \dodoi{10.1088/0004-637X/772/2/93}

\bibitem[{Kušmić(2024)}]{kusmic_2024_pycosie}
Kušmić, S. 2024, {pycosie - Python analysis software for Technicolor Dawn}, 0.1.5,  Zenodo, \dodoi{10.5281/zenodo.12192605}

\bibitem[{{Langan} {et~al.}(2020){Langan}, {Ceverino}, \& {Finlator}}]{Langan:2020}
{Langan}, I., {Ceverino}, D., \& {Finlator}, K. 2020, \mnras, 494, 1988, \dodoi{10.1093/mnras/staa880}

\bibitem[{{Madau} \& {Dickinson}(2014)}]{Madau:2014}
{Madau}, P., \& {Dickinson}, M. 2014, \araa, 52, 415, \dodoi{10.1146/annurev-astro-081811-125615}

\bibitem[{{Nakajima} {et~al.}(2023){Nakajima}, {Ouchi}, {Isobe}, {Harikane}, {Zhang}, {Ono}, {Umeda}, \& {Oguri}}]{Nakajima:2023}
{Nakajima}, K., {Ouchi}, M., {Isobe}, Y., {et~al.} 2023, \apjs, 269, 33, \dodoi{10.3847/1538-4365/acd556}

\bibitem[{{P{\'e}rez-Montero} {et~al.}(2021){P{\'e}rez-Montero}, {Amor{\'\i}n}, {S{\'a}nchez Almeida}, {V{\'\i}lchez}, {Garc{\'\i}a-Benito}, \& {Kehrig}}]{Perez-Montero:2021}
{P{\'e}rez-Montero}, E., {Amor{\'\i}n}, R., {S{\'a}nchez Almeida}, J., {et~al.} 2021, \mnras, 504, 1237, \dodoi{10.1093/mnras/stab862}

\bibitem[{{Price} {et~al.}(2024){Price}, {Bezanson}, {Labbe}, {Furtak}, {de Graaff}, {Greene}, {Kokorev}, {Setton}, {Suess}, {Brammer}, {Cutler}, {Leja}, {Pan}, {Wang}, {Weaver}, {Whitaker}, {Atek}, {Burgasser}, {Chemerynska}, {Dayal}, {Feldmann}, {F{\"o}rster Schreiber}, {Fudamoto}, {Fujimoto}, {Glazebrook}, {Goulding}, {Khullar}, {Kriek}, {Marchesini}, {Maseda}, {Miller}, {Muzzin}, {Nanayakkara}, {Nelson}, {Oesch}, {Shipley}, {Smit}, {Taylor}, {van Dokkum}, {Williams}, \& {Zitrin}}]{Price:2024}
{Price}, S.~H., {Bezanson}, R., {Labbe}, I., {et~al.} 2024, arXiv e-prints, arXiv:2408.03920, \dodoi{10.48550/arXiv.2408.03920}

\bibitem[{{Sanders} {et~al.}(2024){Sanders}, {Shapley}, {Topping}, {Reddy}, \& {Brammer}}]{Sanders:2024}
{Sanders}, R.~L., {Shapley}, A.~E., {Topping}, M.~W., {Reddy}, N.~A., \& {Brammer}, G.~B. 2024, \apj, 962, 24, \dodoi{10.3847/1538-4357/ad15fc}

\bibitem[{{Santini} {et~al.}(2023){Santini}, {Fontana}, {Castellano}, {Leethochawalit}, {Trenti}, {Treu}, {Belfiori}, {Birrer}, {Bonchi}, {Merlin}, {Mason}, {Morishita}, {Nonino}, {Paris}, {Polenta}, {Rosati}, {Yang}, {Boyett}, {Bradac}, {Calabr{\`o}}, {Dressler}, {Glazebrook}, {Marchesini}, {Mascia}, {Nanayakkara}, {Pentericci}, {Roberts-Borsani}, {Scarlata}, {Vulcani}, \& {Wang}}]{Santini:2023}
{Santini}, P., {Fontana}, A., {Castellano}, M., {et~al.} 2023, \apjl, 942, L27, \dodoi{10.3847/2041-8213/ac9586}

\bibitem[{{Seeyave} {et~al.}(2023){Seeyave}, {Wilkins}, {Kuusisto}, {Lovell}, {Irodotou}, {Simmonds}, {Vijayan}, {Thomas}, {Roper}, {Byrne}, {Jones}, {Turner}, \& {Conselice}}]{Seeyave:2023}
{Seeyave}, L. T.~C., {Wilkins}, S.~M., {Kuusisto}, J.~K., {et~al.} 2023, \mnras, 525, 2422, \dodoi{10.1093/mnras/stad2487}

\bibitem[{{Stanway} \& {Eldridge}(2018)}]{Stanway:2018}
{Stanway}, E.~R., \& {Eldridge}, J.~J. 2018, \mnras, 479, 75, \dodoi{10.1093/mnras/sty1353}

\bibitem[{{Stark}(2016)}]{Stark:2016}
{Stark}, D.~P. 2016, \araa, 54, 761, \dodoi{10.1146/annurev-astro-081915-023417}

\bibitem[{{Stefanon} {et~al.}(2021){Stefanon}, {Bouwens}, {Labb{\'e}}, {Illingworth}, {Gonzalez}, \& {Oesch}}]{Stefanon:2021}
{Stefanon}, M., {Bouwens}, R.~J., {Labb{\'e}}, I., {et~al.} 2021, \apj, 922, 29, \dodoi{10.3847/1538-4357/ac1bb6}

\bibitem[{{Sun} \& {Furlanetto}(2016)}]{Sun:2016}
{Sun}, G., \& {Furlanetto}, S.~R. 2016, \mnras, 460, 417, \dodoi{10.1093/mnras/stw980}

\bibitem[{{Tang} {et~al.}(2023){Tang}, {Stark}, {Chen}, {Mason}, {Topping}, {Endsley}, {Senchyna}, {Plat}, {Lu}, {Whitler}, {Robertson}, \& {Charlot}}]{Tang:2023}
{Tang}, M., {Stark}, D.~P., {Chen}, Z., {et~al.} 2023, \mnras, 526, 1657, \dodoi{10.1093/mnras/stad2763}

\bibitem[{{Trump} {et~al.}(2023){Trump}, {Arrabal Haro}, {Simons}, {Backhaus}, {Amor{\'\i}n}, {Dickinson}, {Fern{\'a}ndez}, {Papovich}, {Nicholls}, {Kewley}, {Brunker}, {Salzer}, {Wilkins}, {Almaini}, {Bagley}, {Berg}, {Bhatawdekar}, {Bisigello}, {Buat}, {Burgarella}, {Calabr{\`o}}, {Casey}, {Ciesla}, {Cleri}, {Cole}, {Cooper}, {Cooray}, {Costantin}, {Croton}, {Ferguson}, {Finkelstein}, {Fujimoto}, {Gardner}, {Gawiser}, {Giavalisco}, {Grazian}, {Grogin}, {Hathi}, {Hirschmann}, {Holwerda}, {Huertas-Company}, {Hutchison}, {Jogee}, {Juneau}, {Jung}, {Kartaltepe}, {Kirkpatrick}, {Kocevski}, {Koekemoer}, {Lotz}, {Lucas}, {Magnelli}, {Matharu}, {P{\'e}rez-Gonz{\'a}lez}, {Pirzkal}, {Rafelski}, {Rose}, {Seill{\'e}}, {Somerville}, {Straughn}, {Tacchella}, {Vanderhoof}, {Weiner}, {Wuyts}, {Yung}, \& {Zavala}}]{Trump:2023}
{Trump}, J.~R., {Arrabal Haro}, P., {Simons}, R.~C., {et~al.} 2023, \apj, 945, 35, \dodoi{10.3847/1538-4357/acba8a}

\bibitem[{{Tumlinson} {et~al.}(2017){Tumlinson}, {Peeples}, \& {Werk}}]{Tumlinson:2017}
{Tumlinson}, J., {Peeples}, M.~S., \& {Werk}, J.~K. 2017, \araa, 55, 389, \dodoi{10.1146/annurev-astro-091916-055240}

\bibitem[{{Turk} {et~al.}(2011){Turk}, {Smith}, {Oishi}, {Skory}, {Skillman}, {Abel}, \& {Norman}}]{Turk:2011}
{Turk}, M.~J., {Smith}, B.~D., {Oishi}, J.~S., {et~al.} 2011, \apjs, 192, 9, \dodoi{10.1088/0067-0049/192/1/9}

\bibitem[{{Vijayan} {et~al.}(2019){Vijayan}, {Clay}, {Thomas}, {Yates}, {Wilkins}, \& {Henriques}}]{Vijayan:2019}
{Vijayan}, A.~P., {Clay}, S.~J., {Thomas}, P.~A., {et~al.} 2019, \mnras, 489, 4072, \dodoi{10.1093/mnras/stz1948}

\bibitem[{{Virtanen} {et~al.}(2020){Virtanen}, {Gommers}, {Oliphant}, {Haberland}, {Reddy}, {Cournapeau}, {Burovski}, {Peterson}, {Weckesser}, {Bright}, {van der Walt}, {Brett}, {Wilson}, {Millman}, {Mayorov}, {Nelson}, {Jones}, {Kern}, {Larson}, {Carey}, {Polat}, {Feng}, {Moore}, {VanderPlas}, {Laxalde}, {Perktold}, {Cimrman}, {Henriksen}, {Quintero}, {Harris}, {Archibald}, {Ribeiro}, {Pedregosa}, {van Mulbregt}, \& {SciPy 1. 0 Contributors}}]{Virtanen:2020}
{Virtanen}, P., {Gommers}, R., {Oliphant}, T.~E., {et~al.} 2020, Nature Methods, 17, 261, \dodoi{10.1038/s41592-019-0686-2}

\end{thebibliography}
\bibliographystyle{aasjournal}

\end{document}